\documentclass[twocolumn,showpacs,preprintnumbers,amsmath,amssymb]{revtex4}
\usepackage{graphicx}

\usepackage{amssymb}
\begin{document}
%\preprint{APS/123-QED}
\draft

\title{Nonequilibrium dynamics of polymer translocation and straightening}
\author{Takahiro Sakaue}
 
\email{sakaue@scphys.kyoto-u.ac.jp}
\affiliation{
Department of Physics, Graduate School of Science, Kyoto University, Kyoto 606-8502, Japan
}%

\date{\today}% It is always \today, today,

\begin{abstract}
When a flexible polymer is sucked into a localized small hole, the chain can initially respond only locally and the sequential nonequilibrium processes follow in line with the propagation of the tensile force along the chain backbone.
We analyze this dynamical process by taking the nonuniform stretching of the polymer into account both with and without hydrodynamics interactions.
Earlier conjectures on the absorption time are criticized and new formulae are proposed together with time evolutions of relevant dynamical variables.
\end{abstract}


\maketitle

\section{Introduction}
\label{Introduction}
A long flexible polymer is one of the representative examples of soft matter.
A common feature of soft matter is a presence of mesoscopic length scales, which is, in many respects, responsible for their unique properties such as the high susceptibility.
For dilute polymer solutions (with each chain made of $N_0$ succession of monomers of size $b$), this corresponds to the Flory radius $R_0=b N_0^{\nu}$ of individual coils, which serves as a basis for the scaling theory~\cite{deGennes}.
From the estimation of the elastic modulus $\sim k_BT/R_0^3$, one can realize an important consequence that a long chain is readily exposed to significant distortions, such as stretching and compression, by rather weak perturbations.

Although the extension of a polymer in various flow fields or by mechanical stretching have been extensively studied~\cite{deGennes,coil-stretch, Pincus, Chu_Larson, trumpet, stem-flower}, most attentions so far have been focused on equilibrium or steady state properties.
One can also ask the dynamical process from one steady state to the other induced by sudden change of external field~\cite{stiff_dynamics2}, which would be important in relation to recent development of micromanipulation techniques.
For instance, imagine that an initially relaxed polymer is suddenly started to be pulled by its one end (see Appendex~\ref{pulling_end}).
If the force is sufficiently weak ${\tilde f}_{R_0}<1$, a chain as a whole follows at the average velocity $v \simeq f/(\eta_s R_0)$ ($\eta_s$ is the solvent viscosity) with keeping the equilibrium conformation.
(Here and in what follows, we denote the dimensionless force ${\tilde f}_x \equiv fx/(k_BT)$, where $x$ has the dimension of the length.)
For large force ${\tilde f}_{R_0}>1$, however, only a part of the chain can respond immediately, while remaining rear part does not feel the force yet.
As time goes on, the tension propagates along the chain, which alters chain conformation progressively, and the steady state is reached after a characteristic time.
At room temperature, the critical force $f \sim (k_BT)/R_0$ is estimated to be on the order of pico Newton for a flexible chain with $N=100 \sim 1000$, comparable to the usual force range in the single molecule manipulation experiments with atomic force microscopy and optical tweezer. The force generated by molecular motors also falls into this range. This implies a possible importance of such a nonequilibrium response in many biological as well as technological situations.

In the present paper, we illustrate such a nonequilibrium response using an example of polymer absorption or aspiration into a small spot.
Our target here is the dynamical process, in which a polymer is sucked into a localized hole.
This is different from the pulling the chain's one end~\cite{Kantor_Kardar}, but similar in the way how the chain responds to the local force, and, in fact, relevant to the dynamics of polymer translocation through hole~\cite{Kasianowicz, DrivenDNA} and the adsorption process to a small particle.
(In this case, the force $f$ is related to the chemical potential change $\Delta \mu$ due to the absorption via $f = \Delta \mu/b$.)
Although the phenomenon of polymer translocation has been an active research topic in the past decade as a model for biopolymer transport through a pore in membrane, our current understanding for the dynamics, in particular the strongly driven case, is restrictive.
So far, the scaling estimates of the characteristic time $\tau$ for absorption process in immobile solvents have been proposed~\cite{Grosberg_absorption, Kantor_Kardar}.
Grosberg et. al. argued the absorption time for a Rouse chain on the ground that Rouse time $\tau_R = \tau_0 N_0^{2}$ is the solitary relevant time scale ($\tau_0 \simeq \eta_s b^3/(k_BT)$ is a microscopic time scale), and all other relevant parameters appear in the dimensionless combination ${\tilde f}_{R_0}$, thus,
\begin{eqnarray}
\tau=\tau_R \phi({\tilde f}_{R_0})
\label{tau_scaling}
\end{eqnarray}
The scaling function $\phi$ is determined from the requirement that the speed of the process, $N_0/\tau$ must be linear in the applied force, leading to
\begin{eqnarray}
\tau = \frac{\eta_s b^2}{f}N_0^{3/2}
\label{tau_Grosberg}
\end{eqnarray}
This result was interpreted as a sequential straightening of "folds"~\cite{Grosberg_absorption}.
For a chain with excluded volume studied by Kantor and Karder, a similar scaling argument leads to~\cite{Kantor_Kardar} 
\begin{eqnarray}
\tau \sim N_0^{1+\nu}/f
\label{tau_Kantor_Karder}
\end{eqnarray}
where Flory exponent $\nu=1/2$ for a chain in $\theta$ solvent, while in good solvent $\nu \simeq 3/5$ (in space dimension ${\mathcal D}=3$) and $\nu = 3/4$ (in ${\mathcal D}=2$).

%One of the simples examples would be the stretching of an initially relaxed polymer by applying a tensile force on its one end at time $t>0$, which is now possible by utilizing various single molecule manipulation techniques.
%In this Letter, we focus on a slightly different situation, in which a polymer is sucked into a localized small hole.

We tackle this problem through different approach by explicitly considering the dynamics of the tension propagation.
This enables us to unveils the physics behind the nonequilibrium driven absorption and go beyond the previous works by taking the excluded volume effect and/or hydrodynamics interactions into account.
We indeed find that eq. (\ref{tau_Grosberg}), (\ref{tau_Kantor_Karder}) are not generally correct as a consequence of the salient feature of flexible chain that the response to the aspiration or stretching force is {\it nonuniform both in space and time}. At very strong driving, the chain finite extensibility matters, and this leads us to propose three distinguished regimes for the absorption process depending on the degree of forcing.
Interestingly, we shall see that eq. (\ref{tau_Grosberg}), (\ref{tau_Kantor_Karder}) are recovered in the limit of very strong forcing only.
%We indeed verify eq. (\ref{tau_Grosberg}, \ref{tau_Kantor_Karder}) in the limit of large force ${\tilde f}_b>N_0^{\nu}$, but it turns out to be modified for smaller driving force reflecting the salient feature of flexible chain, i.e., response to the aspiration or stretching force is {\it nonlocal both in space and time}.
In addition to the absorption time, we can also predict the time evolutions of dynamical variables governing this nonequilibrium process.
Below, the problem of the dynamical response is formulated with basic equations and the absorption dynamics is analyzed in Sec.~\ref{Dynamical response}. 
Then, Sec.~\ref{inextensibility} and \ref{hydrodynamics} are devoted to discussions on the effect of the finite chain extensibility and hydrodynamic interactions, respectively.
Summary and future perspectives are given in Sec.~\ref{summary}.
Some technical details and mathematics are given in Appendix~\ref{continuity_eq} and \ref{solve}.
Another example of nonequilibrium response, i.e., a sudden pulling of a chain by its one end, is briefly discussed in Appendix~\ref{pulling_end}.

\section{Formulation}
\subsection{Dynamical response to strong forcing}
\label{Dynamical response}
\begin{figure}
\includegraphics[width=7cm]{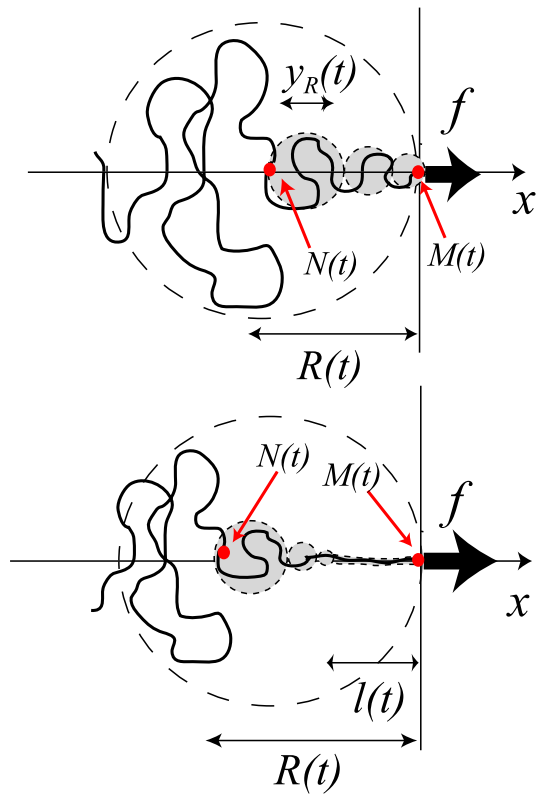}
\caption{A dynamical response of a polymer initially at rest to the strong aspiration. A chain is pulled from ($x<0$) region to the aspiration spot located at the origin. Monomers in gray region are under the influence of the tensile force, while monomers in the rear part is yet relaxed. The distance between the boundary (associated with tension propagation) and the origin and the size of the largest blob at the boundary are denoted as $R(t)$ and $y_R (t)$, respectively. Monomers are sequentially labelled as $1, 2, \cdots N_0$ from one end (arriving at the origin initially) to the other. $M(t)$ and $N(t)$ are, respectively, labels of the monomers residing at the origin and the boundary at time $t$. (top) $N_0^{-\nu}< {\tilde f}_b < 1$; (bottom) $1<{\tilde f}_b < N_0^{\nu}$, in which the anterior part ($-l(t)<x<0$) is completely stretched.}
\label{suction_scheme}
\end{figure}

Now, let us imagine the moment, when one end of the chain arrives at the attractive hole at the origin (Fig. \ref{suction_scheme}).
%Let us first estimate the number of monomers which can respond the applied force immediately.
The fact that the first monomer is strongly pulled influences the rear vicinity immediately, but not far away.
If we notice the subunit consisted of first $g_0$ monomers of size $y_0=b g_0^{\nu}$, this subunit starts to move with the average velocity $v_0 \simeq f/(\eta_s y_0)$, provided that the deformation is insignificant in this scale.
The size of such a subunit is deduced by comparing the longest relaxation time of the subunit $\tau_0 g_0^{3\nu}$ to the velocity gradient $v_0/y_0$
\begin{eqnarray}
g_0 \simeq \left( \frac{k_BT}{fb} \right)^{1/\nu}  \Longleftrightarrow y_0 \simeq \frac{k_BT}{f}
\label{initial}
\end{eqnarray}
and this constitutes the initial condition.

The absorption process proceeds with time; at time $t$, the tension is transmitted up to $N(t)$-th monomer, while $M(t)$ front monomers are already absorbed (Fig. 1).
This indicates that one can regard that the chain portion at ($-R(t) < x < 0$) takes a steady state conformation moving with the average velocity $v(t)$.
%The large tensile force set position dependent length scale $y(x)$, above which the chain is substantially elongated.
Here the drag force builds up along the chain starting from a free boundary ($x=-R(t)$) to the origin, which makes the overall chain shape reminiscent to a ``trumpet"~\cite{trumpet}.
It means that the length scale $y(x)$ set by the large tensile force, above which the chain is substantially elongated, is position dependent, thus, the elastic behaviour of the chain is described as a sequence of blobs with growing size $y(x)=bg(x)^{\nu}$.
%Denoting the lateral size of the chain $y(x)$, the local force balance reads%
This leads to the following local force balance equation;
\begin{eqnarray}
y(x) \simeq \frac{k_BT}{\eta_s v(t) }\left(\frac{1}{x+R(t)}\right) \qquad [-R(t)+y_R(t) \le x \le 0 ] \label{f_balance1}
\end{eqnarray}
where we introduce the cut-off length $y_{R}(t)$, which signifies the size of the largest blob at the free boundary;
\begin{eqnarray}
&&y(-R(t)+y_R(t)) \simeq y_R(t) \nonumber \\
\Longleftrightarrow 
&&y_R(t) \simeq \left( \frac{k_BT}{\eta_s v(t)}\right)^{1/2} \quad[\mbox{at} \ x=-R(t)+y_{R}(t)]
\label{boundary2}
\end{eqnarray}
(Note that throughout the paper, we neglect the logarithmic factor associated with the friction of asymmetric objects in low Reynolds number~\cite{RW_Biology} regime as well as other numerical coefficients of order unity unless specified.)
The mass conservation reads
\begin{eqnarray}
%\int^{0}_{-R(t)+{\tilde y} (t)} \rho(x,\ t) S(x,\ t) dx + \frac{{\tilde g} (t)}{2} + M(t)= N(t)
\int^{0}_{-R(t)} \rho(x) S(x) dx  + M(t)&=& N(t) \label{mass_cons1}
\end{eqnarray}
where $\rho(x) \simeq g(x)/y(x)^3$  is the monomer density ($y(x)= b g(x)^{\nu}$) and $S(x) \simeq y(x)^2$ is the cross-sectional area of the conformation.
Equation (\ref{f_balance1}) and (\ref{mass_cons1}) constitutes basic equations supplemented with the following statistical relation available from the conformation at rest ($t<0$)
\begin{eqnarray}
b N(t)^{\nu} = R(t) 
\label{R_N_statistical}
\end{eqnarray}
and ``boundary conditions" both at the free boundary eq. (\ref{boundary2}) and the origin
\begin{eqnarray}
\eta_s R(t) v(t) \simeq f
\label{boundary1}
\end{eqnarray}
These conditions express the force balance at the free end and the total force balance, respectively.
%\begin{eqnarray}
%y(0,\ t) \simeq \frac{k_BT}{f} \Longleftrightarrow R(t) \simeq \frac{f}{\eta v(t)}  \quad [\mbox{at} \ x=0] \label{boundary1}
%\end{eqnarray}

After casting Eq. (\ref{mass_cons1}) in the differential form
\begin{eqnarray}
%\rho(-R(t) + {\tilde y} (t))S(-R(t)+ {\tilde y} (t)) \left( \frac{dR(t)}{dt} + v(t) \right) = \frac{dN(R)}{dt}
%\rho|_{x=-R(t)}S|_{x=-R(t)} \left( \frac{dR(t)}{dt} + v(t) \right) = \frac{dN(R)}{dt}
[\rho S]_{(x=-R(t))} \left( \frac{dR(t)}{dt} + v(t) \right) = \frac{dN(R)}{dt},
\label{mass_cons2}
\end{eqnarray}
we obtain the following equation for the tension propagation (see Appendix);
\begin{eqnarray}
%({\tilde f}_{R(t)})^{\frac{1-3\nu}{2\nu}}\left(\frac{R(t)}{b} \right)^{3}\left[ 1-B_0 ({\tilde f}_{R(t)})^{\frac{\nu-1}{2\nu}}\right] = \frac{t}{\tau_0}
\label{R_t_1}
\left(\frac{R(t)}{R_0}\right)^{\frac{3\nu+1}{2\nu}}\frac{\Phi(R(t))}{\Phi(R_0)} = \frac{t}{\tau_1}
\label{R_t}
\end{eqnarray}
With the symbol $x$ possessing the dimension of length (recall once again the definition of the dimensionless force ${\tilde f}_x$), the function $\Phi(x)$ is defined as
\begin{eqnarray}
\Phi(x) &\equiv& 1-B_0({\tilde f}_x)^{\frac{\nu-1}{2\nu}}+(B_0-1)({\tilde f}_x)^{-\frac{3\nu+1}{2\nu}} \label{Phi_definition}\\
&\simeq&1-B_0({\tilde f}_x)^{\frac{\nu-1}{2\nu}}
\end{eqnarray}
where $B_0$ is the numerical coefficient of order unity and $\tau_1$ is the time, when the other side of the chain end reaches the steady state and gets set into motion due to the tensile force; $R(\tau_1) = R_0$, thus,
\begin{eqnarray}
\tau_1 = \tau_Z ({\tilde f}_{R_0})^{\frac{1-3\nu}{2\nu}} \Phi(R_0)
\label{tau_1}
%\tau_1 = \tau_0  ({\tilde f}_b)^{\frac{1-3\nu}{2\nu}} N_0^{\frac{3\nu+1}{2}}\left[1 - B_0 ({\tilde f}_b)^{\frac{\nu-1}{2\nu}} N_0^{\frac{\nu-1}{2}}\right]
%&\simeq& \tau_0  ({\tilde f}_b)^{\frac{1-3\nu}{2\nu}} N_0^{\frac{3\nu+1}{2}}
\end{eqnarray}
where $\tau_Z = \tau_0 N_0^{3\nu}$ is the Zimm time.
We notice that the time $\tau_1$ obtained here satisfies the scaling form of eq. (\ref{tau_scaling}) with the replacement of $\tau_R$ by $\tau_Z$.

The number of the monomers absorbed is calculated as
\begin{eqnarray}
M(t) = \int^t_0 [\rho S]_{x=0} \ v(s) ds 
\label{eq_for_M}
\end{eqnarray}
which can be shown to reach $M_1 = N_0(1-({\tilde f}_{R_0})^{(\nu-1)/\nu})$ at $t=\tau_1$.
After that, the whole yet non-absorbed part of chain is under the tension and pulled toward the hole, thus, the evolution of $M(t)$ is governed by
\begin{eqnarray}
-\eta_s \left( \frac{dL(t)}{dt}\right) L(t) \simeq f
\label{diff_eq_after}
\end{eqnarray}
where $L(t)$ is the long axis length of the non-absorbed chain.
This leads to~\footnote{The time evolution of $L(t)$ is easily derived by integrating eq.~(\ref{diff_eq_after}) from $t=\tau_1$ to $t$ by noting $L(\tau_1)=R_0$. Then, by substituting the velocity $v(t) \simeq f/(\eta_s L(t))$ into eq.~(\ref{eq_for_M}), one obtains eq.~(\ref{M_after}) .}
\begin{eqnarray}
M(t) = M_1 + {\tilde f}_{R_0} ({\tilde f}_{b})^{-\frac{1}{\nu}}\left[ 1-\left\{ 1-{\tilde f}_{R_0} \left(\frac{t-\tau_1}{\tau_Z}\right)\right\}^{\frac{1}{2}}\right] \label{M_after}\\
\quad [t > \tau_1] \nonumber
\end{eqnarray}
The absorption process completes at time $\tau \equiv \tau_1 + \tau_2$ with
\begin{eqnarray}
\tau_2 \simeq \tau_Z \ ({\tilde f}_{R_0})^{-1}
\end{eqnarray}

\begin{figure}
\includegraphics[width=6.5cm]{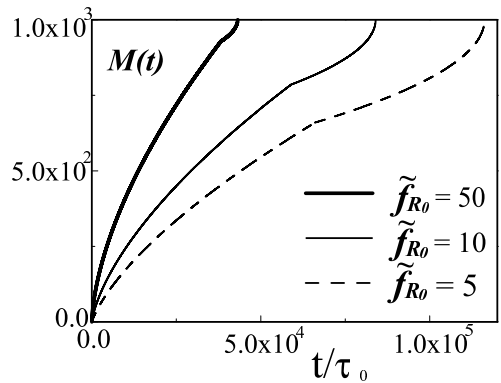}
\caption{Time evolutions of number of absorbed monomers $M(t)$ at various forces for the chain length $N_0=10^3$ under good solvent condition ($\nu=3/5$). Inflection-like points mark the end of the tension propagation stage ($t=\tau_1$ and $M=M_1$).}
\label{evolution_M}
\end{figure}
\begin{figure}
\includegraphics[width=6.5cm]{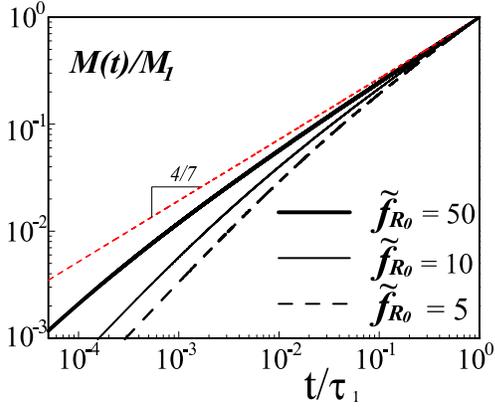}
\caption{Normalized time evolutions (double logarithmic scale) of number of absorbed monomers $M(t/\tau_1)/M_1$ at various scaled forces in the tension propagation stage under good solvent condition ($\nu=3/5$).}
\label{evolution_M_normalize}
\end{figure}

The time evolution of $M(t)$ is plotted in Fig.~\ref{evolution_M} for $N_0=10^3$ under various forces.
One can see that the first stage of the tension propagation dominates the most process under the large force ${\tilde f}_{R_0} >1$.
In this case, from eq.~(\ref{tau_1}), the absorption time is approximated as
\begin{eqnarray}
\tau \simeq \tau_Z \ ({\tilde f}_{R_0})^{\frac{1-3\nu}{2\nu}} \simeq \tau_0  \ ({\tilde f}_b)^{\frac{1-3\nu}{2\nu}} N_0^{\frac{3\nu+1}{2}}
\label{tau_trumpet}
\end{eqnarray}
In more specific form, $\tau \sim N^{7/5}/f^{2/3}$ for a chain in good solvent ($\nu=3/5$) and $\tau \sim N^{5/4}/f^{1/2}$ for a chain in $\theta$ solvent, which are apparently different from earlier conjectures (eq. (\ref{tau_Grosberg}), (\ref{tau_Kantor_Karder})).
As eq.~(\ref{R_t}) indicates, time evolutions of dynamical variables with suitable normalization are represented by master curves parametrized by values of ${\tilde f}_{R_0}$ in this tension propagation stage (thus, almost all the absorption process).
 Figure~\ref{evolution_M_normalize} exemplifies the time evolution of $M(t)$ with the normalization $M(t)/M_1$ and $t/\tau_1$, where $M_1 = N_0(1-({\tilde f}_{R_0})^{(\nu-1)/\nu})$.
%and almost the whole evolution of $M(t)$ are represented by master curves parametrized by values of ${\tilde f}_{R_0}$ with the normalization $M(t)/M_1$ and $t/\tau_1$, where $M_1 = N_0(1-({\tilde f}_{R_0})^{(\nu-1)/\nu}$ (Fig.~\ref{evolution_M_normalize}).
The larger the scaled force ${\tilde f}_{R_0}$, the sooner the master curve approaches the asymptotic line with the slope $(1+\nu)/(1+3\nu)$.

\subsection{Effect of the finite chain extensibility}
\label{inextensibility}
The above argument is valid as long as ${\tilde f}_b<1$.
For larger forces, as one can see from eq. (\ref{initial}), the chain close to the origin is completely stretched (see the bottom of Fig.~\ref{suction_scheme}).
To include such an effect, let us begin with the limit of strong force (${\tilde f}_b > N_0^{\nu}$)\footnote{This limit was briefly discussed by D. Lubensky in a chapter of ref.~\cite{Kasianowicz}.}.
Then, at any moment, the entropic coiling is completely irrelevant for the tensed part of the chain.
We can repeat the same argument by setting $y(x) = b$ for ($-R(t) < x < 0$), and find
\begin{eqnarray}
\left( \frac{R(t)}{R_0}\right) = \left( \frac{t}{\tau_1}\right)^{\frac{\nu}{1+\nu}}
\label{R_t_stronglimit}
\end{eqnarray}
with
\begin{eqnarray}
\tau_1 = \tau_0 \ ({\tilde f}_b)^{-1} N_0^{1+\nu} = \tau_R \ ({\tilde f}_{R_0})^{-1}
\label{tau_1_strong_pull}
\end{eqnarray}
where Rouse time $\tau_R = \tau_0 N_0 (R_0/b)^2$ (for a chain with Flory exponent $\nu$) appears as a characteristic time reflecting the fact that the backflow effect is nearly negligible in this strong pulling limit.
It is noted that eq. (\ref{tau_1_strong_pull}) coincides with eq. (\ref{tau_Grosberg}) and (\ref{tau_Kantor_Karder}).

For the intermediate case of $1< {\tilde f}_b < N_0^{\nu}$, this process of strong pulling limit is valid until the time $\tau_1^*$.
This time is determined by monitoring the monomer at the free boundary; the velocity of the tense string of length $l(t)=N(t)b$ is given by
\begin{eqnarray}
v(t) = b \frac{d N(R)}{dt} \simeq b \left(\frac{{\tilde f}_b}{\tau_0}\right)^{\frac{1}{1+\nu}}t^{-\frac{\nu}{1+\nu}}
\end{eqnarray}
where we have utilized eq.~(\ref{R_N_statistical}) and (\ref{R_t_stronglimit}).
This quantity decreases with time and at $t=\tau_1^*$, the drag force for the free boundary monomer becomes comparable to $k_BT/b$.
\begin{eqnarray}
\eta_s b v(\tau_1^*) \simeq \frac{k_BT}{b}
\end{eqnarray}
From this, we find
\begin{eqnarray}
\tau_1^* = \tau_0 \ ({\tilde f}_b)^{\frac{1}{\nu}}
\label{tau1*}
\end{eqnarray}
After $\tau_1^*$, the conformation of the chain under tension is a completely stretched string of length $l(t)$ followed by coiled subunits with growing size~\cite{stem-flower}.
For the latter part, we can apply the aforementioned analysis for the trumpet, where the ``boundary condition" is imposed not at the origin but at $x=l(t)$.
Setting $R(t)=l(t)$ and $y(t)=b$ in the local force balance equation (eq.~(\ref{f_balance1})) yields
\begin{eqnarray}
l(t) \simeq R(t) -\frac{k_BT}{\eta_s b v(t)} \simeq R(t) ( 1- ({\tilde f}_b)^{-1})
\label{l_R}
\end{eqnarray}
where the second equality utilizes the total force balance $\eta_s v(t) R(t) \simeq f$.
More precisely, eq. (\ref{l_R}) should be written as
\begin{eqnarray}
p \equiv \frac{l(t)}{R(t)} \simeq 1-c_0 \ ({\tilde f}_b)^{-1}
\label{define_p}
\end{eqnarray}
where $c_0$ is the logarithmic factor dependent on $l(t)$ and $R(t)$.
Ignoring this small correction allows us to derive the following form of the tension propagation dynamics at $t>\tau_1^*$
\begin{eqnarray}
\frac{t-\tau_1^*}{\tau_1 -\tau_1^*}= \left( \frac{R(t)}{R_0}\right)^{\frac{3\nu+1}{2\nu}} \frac{\Phi^*(R(t))}{\Phi^*(R_0)}
\label{R_t_stem_flower}
\end{eqnarray}
where the function $\Phi^*(x)$ and the time $\tau_1^*$ are respectively defined as
\begin{eqnarray}
\Phi^* (x) &\equiv& \left( 1-p^{\frac{3\nu+1}{2\nu}}\right) -B_0({\tilde f}_{x})^{\frac{\nu-1}{2\nu}}\left(1-p^2 \right) \\
\tau_1-\tau_1^* &=& \tau_Z \ ({\tilde f}_{R_0})^{\frac{1-3\nu}{2\nu}}\Phi^*(R_0)
\label{tau1-tau1*}
\end{eqnarray}
Absorption time is approximated by the time $\tau_1$ when the tension reaches the other end, thus, from eq. (\ref{tau1*}) and (\ref{tau1-tau1*})
\begin{eqnarray}
\tau \simeq \tau_Z ({\tilde f}_{R_0})^{\frac{1-3\nu}{2\nu}}\ \left[ \Phi^*(R_0) + \left( \frac{{\tilde f}_b}{N_0^{\nu}}\right)^{\frac{1+3\nu}{2\nu}}\right]
\label{tau_stem_flower}
\end{eqnarray}
Utilizing the asymptotic behaviours of eq. (\ref{define_p}), i.e., $p \rightarrow 0$ for ${\tilde f}_b \rightarrow 1$, and $p \rightarrow 1$ for ${\tilde f}_b >>1$, eq. (\ref{tau_stem_flower}) shows smooth crossovers to the trumpet regime eq. (\ref{tau_trumpet}) at ${\tilde f}_b \rightarrow 1$ and strong pulling limit eq. (\ref{tau_1_strong_pull}) at ${\tilde f}_b \rightarrow N_0^{\nu}$.

\subsection{Effect of hydrodynamic interactions}
\label{hydrodynamics}
At certain situations, hydrodynamic interactions may become screened, i.e., a polymer confined in narrow slit or in melt of short chains.
Then, a question arises; what is the effect of the hydrodynamics in this dynamical process?
To answer this, let us "switch off" the induced flow of solvents.
Then, the only requisite alteration is the dissipation mechanism, which becomes local and independent of the chain conformation.
The local force balance is, instead of eq. (\ref{f_balance1}),  written as
\begin{eqnarray}
\frac{k_BT}{y(x)} \simeq \eta_s b v(t) \int_{-R(t)}^x dx' \ \frac{g(x')}{y(x')}
\end{eqnarray}
where $g(x)$ is related to $y(x)$ through $y(x) = b g(x)^{\nu}$.
The largest blob size at the free boundary is
\begin{eqnarray}
y_{R(t)} \simeq b \left( \frac{k_BT}{b^2 \eta_s v(t)}\right)^{\frac{\nu}{1+\nu}}
\end{eqnarray}
One can analyze along the same line as the chain in mobile solvents.
In partucular, $\tau_1$ for the moderate forcing (${\tilde f}_b<1$) is obtained as
\begin{eqnarray}
\tau_1 = \tau_{R} \ ({\tilde f}_{R_0})^{-\frac{2}{1+\nu}} \Phi_R(R_0)
\end{eqnarray}
with
\begin{eqnarray}
\Phi_R(x) &\equiv& 1-B_0({\tilde f}_x)^{\frac{\nu-1}{\nu(\nu+1)}}+(B_0-1)({\tilde f}_x)^{-\frac{2\nu^2+\nu+1}{\nu(\nu+1)}}\\
&\simeq& 1-B_0({\tilde f}_x)^{\frac{\nu-1}{\nu(\nu+1)}}
\end{eqnarray}
which coincides with the scaling form of eq. (\ref{tau_scaling}) and should be contrasted with eq. (\ref{tau_1}) with hydrodynamic interactions.
On the other hand, the result in the limit of the strong forcing (${\tilde f}_b>N_0^{\nu}$) is not altered, and $\tau_1$ is given by eq. (\ref{tau_1_strong_pull}).
It is in this limit only that the requirement $\tau \sim f^{-1}$ is fulfilled reflecting the saturation of chain deformation due to the complete stretching, therefore, the earlier conjectures are approved~\cite{Grosberg_absorption, Kantor_Kardar}.
For smaller forces, the soft elasticity of the chain results in more involved responses as we have seen; the stronger driving results in the more intense chain deformation, and this deformation behaviour affects the absorption dynamics.

\section{Summary and perspectives}
\label{summary}
There would be many practical situations, in which the externally imposed velocity gradient exceeds the inverse relaxation time of long polymers.
If the external field acts locally, effects associated with nonequilibrium response, i.e., the propagation of the tensile force along the chain, are expected to show up.
We focused on the problem of absorption or aspiration into a localized hole and demonstrated how such a process can be physically described.

One of the most relevant situations of aspiration dynamics studied here is the polymer translocation through a pore.
We should note, however, in the problem of biopolymer transport through a membrane pore, the role of specific interactions may become essentially important~\cite{Kasianowicz,DrivenDNA,Lubensky_Neslon}.
We neglect all the complications associated with such a factor and focused on universal aspects as a consequence of a polymeric nature, in this sense, the present analysis may be regarded as an ``ideal" version in view of the relation with the problem of polymer translocation.
Such an ``ideal" situation would be now experimentally feasible thanks to the advance in nanoscale fabrications~\cite{Biance}.

The advantage of the present framework is rather wide range of applicability to related problems. As mentioned in Sec.~\ref{Introduction}, if the chain is suddenly pulled its one end, the response is {\it nonuniform} both in space and time. The resultant transient dynamics of the chain extension can be analyzed in a similar way (see Appendex~\ref{pulling_end}). The same physics is also expected in the escape process of a confined polymer from a planner slit~\cite{Confinement_driven}.
Considering hierarchical structures common in polymeric systems, such nonequilibrium dynamics in a single chain level would be expected to show up in macroscopic material properties, too.
We hope that the present study provides the basic insight involved in the driven nonequilibrium process of polymer absorption and its related problems and future investigations including the comparison with the numerical~\footnote{Simulation studies so far were conducted with large force and numerical data with weaker forcing (${\tilde f}_{b} <1$) seems to be currently unavailable. Therefore, seeking for distinguishable regimes depending on the magnitude of the force as well as the effect of hydrodynamic interactions remain to be seen. A care should be taken for modelling the chain connectivity to ensure the finite chain extensibility, otherwise the backbone of the model chain would be overstretched and becomes unrealistic under large forcing (${\tilde f}_b > 1$). } and even real experiments would be valuable.

\begin{acknowledgments}
I wish to thank T. Ohta for fruitful discussions. This research was supported by JSPS Research Fellowships for Young Scientists (No. 01263).
\end{acknowledgments}

\appendix
\section{Continuity equation}
\label{continuity_eq}
In this appendix, we shall discuss the relation between integral and differential forms of the mass conservation equation.
We define $h(x) = \rho(x) S(x)$ and $\int h(x) dx = H(x)$ for the concise notation.
Then, the integral form of the mass conservation (eq.~(\ref{mass_cons1})) is written as
\begin{eqnarray}
H(0)-H(-R(t))+M(t)=N(t)
\end{eqnarray}
The variation of this equation with the variable transformation from $x$ to $u=x-v(t)t$ leads to
\begin{eqnarray}
&&dH\bigr|_{u=-v(t)t}-dH \bigr|_{u=-R(t)-v(t)t}+dM(t) =dN(t) \\
\Leftrightarrow
&&h(0)\times[-vdt] -h(-R(t))\times[d(-R(t))-v(t)dt]+dM(t) \nonumber \\
&&= \frac{dN}{dR}dR(t)
\end{eqnarray}
Since $h(0) \times v(t)dt$ is the number of absorbed monomer during the time interval $dt$, the first and the last terms in left hand side are canceled out.
The resultant equation is the differential form of the mass conservation (eq.~(\ref{mass_cons2})).

\section{Solving differential equations}
\label{solve}
In this appendix, we illustrate some technical details how to solve the differential equations.
Let us see eq.~(\ref{mass_cons2}), which is coupled with eq.~(\ref{boundary2}), (\ref{R_N_statistical}) and (\ref{boundary1}).
Using the definition of density $\rho(x)$ and the cross-sectional area $S(x)$ (given below eq.~(\ref{mass_cons1})) and also eq.~(\ref{R_N_statistical}), one obtains
\begin{eqnarray}
\left( R(t)^{\frac{1-\nu}{\nu}} - (y_{R(t)})^{\frac{1-\nu}{\nu}} \right) \frac{dR(t)}{dt} \simeq (y_{R(t)})^{\frac{1-\nu}{\nu}} v(t)
\label{diff_1}
\end{eqnarray}
By putting force balance conditions (eq.~(\ref{boundary2}) and (\ref{boundary1})), eq.~(\ref{diff_1}) is transformed to
\begin{eqnarray}
\left[R(t)^{\frac{1-\nu}{\nu}} - \left(\frac{k_B T}{f} R(t) \right)^{\frac{1-\nu}{2\nu}} \right] \frac{dR(t)}{dt} \nonumber \\
\simeq \left( \frac{k_B T}{f} R(t) \right)^{\frac{1-\nu}{2\nu}} \frac{f}{\eta_s R(t)}
\label{diff_2}
\end{eqnarray}
The solution of eq.~(\ref{diff_2}) with the initial condition eq.~(\ref{initial}) is
\begin{eqnarray}
({\tilde f}_{b})^{\frac{1-3\nu}{2\nu}} \left[ \left( \frac{R(t)}{b}\right)^{\frac{3\nu+1}{2\nu}} \right]^{R(t)}_{y_0} \nonumber \\
 -B_0({\tilde f}_{b})^{-1} \left[\left( \frac{R(t)}{b}\right)^2 \right]^{R(t)}_{y_0}  =\frac{t}{\tau_0}
\label{diff_3}
\end{eqnarray}
where we introduce the numerical coefficient $B_0$ of order unity to replace the relation symbol $\simeq$ with $=$.
Using the function $\Phi$ (eq.~(\ref{Phi_definition})), the above equation is rewrriten in the compact form
\begin{eqnarray}
({\tilde f}_{R(t)})^{\frac{1-3\nu}{2\nu}}\left( \frac{R(t)}{b}\right)^3 \Phi(R(t))= \frac{t}{\tau_0}
\end{eqnarray}
The time $\tau_1$ (eq.~(\ref{tau_1})), when the tensile force reaches the chain end is obtained by putting $R(t)=R_0$ in this equation.
Equation~(\ref{R_t}) is obtained after normalizing $t$ by $\tau_1$.
Equations~(\ref{R_t_stem_flower}) and (\ref{tau_stem_flower}) in the case of strong pulling ($1< {\tilde f}_b < N_0^{\nu}$) are obtained in the same way just by replacing the lower limit of the integral in eq.~(\ref{diff_3}) with $l(t)$ (eq.~(\ref{l_R})) and the right hand side with $(t-\tau_1^*)/\tau_0$.

\section{Pulling one end}
\label{pulling_end}
\begin{figure}
\includegraphics[width=7cm]{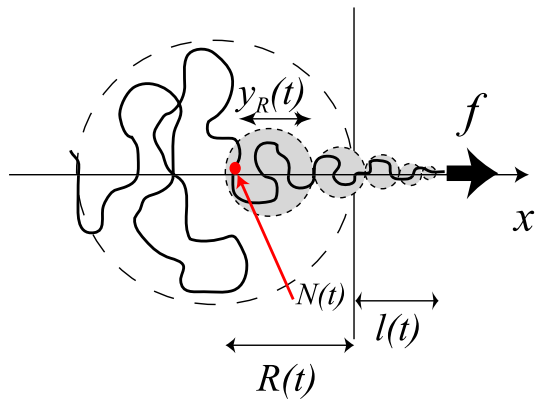}
\caption{A dynamical response of a polymer initially at rest to the strong pulling by its one end ($N_0^{-\nu}< {\tilde f}_b < 1$). A coordinate is defined in such a way that the origin is the position of one chain end at time $t=0$ and that end is started to be pulled to right ($x>0$) direction. Monomers in gray region are under the influence of the tensile force, while monomers in the rear part is yet relaxed. The distance between the boundary (front of the tension propagation) and the origin and the size of the largest blob at the boundary are denoted as $R(t)$ and $y_R (t)$, respectively. Monomers are sequentially labelled as $1, 2, \cdots N_0$ from the pulled end to the other. $N(t)$ is the label of the monomer residing at the boundary at time $t$.}
\label{pull_scheme}
\end{figure}

Here, we shall briefly discuss another example of nonequilibrium response, i.e., the transient dynamics of the chain stretching pulled by its one end (Fig.~\ref{pull_scheme}).
The difference between the absorption and the present stretching process was pointed out by Kantor and Kardar~\cite{Kantor_Kardar}, which is clearly recognized by comparing Fig.~\ref{suction_scheme} and Fig.~\ref{pull_scheme}; in the former, the site of action is fixed in space and the chain portion after crossing this point gets relaxed, while in the latter case, the site of action is moving with the velocity $v(t)=dl(t)/dt$ and the chain portion after crossing the origin is still tensed and contributes the friction.

In this case, too, one can distinguish three different regimes depending on the applied force; (i) trumpet (${\tilde f}_b <1<{\tilde f}_{R_0}$), (ii) intermediate ($1<{\tilde f}_b <N_0$) and (iii) strong limit with complete stretching (${\tilde f}_b>N_0$) by monitoring the force acting on the last monomer at the free boundary (Note that the border between regime (ii) and (iii) is different from the absorption case).
Just as the case in the absorption dynamics, one can write down basic equations.
The local force balance;
\begin{eqnarray}
y(x) \simeq \frac{k_BT}{\eta_s v(t) }\left(\frac{1}{x+R(t)}\right)  \ [-R(t)+y_R(t) \le x \le l(t) ] \label{f_balance_end}
\end{eqnarray}
The mass conservation;
\begin{eqnarray}
\int^{l(t)}_{-R(t)} \rho(x) S(x) dx = N(t) \label{mass_cons_end}
\end{eqnarray}
One also needs eq.~(\ref{R_N_statistical}), the condition for the free boundary (eq.~(\ref{boundary2})) and the total force balance;
\begin{eqnarray}
y(x=l(t))\simeq \frac{k_BT}{f}\\
\Leftrightarrow \ \ l(t)+R(t) \simeq \frac{f}{\eta_s v(t)}
\label{boundary_end}
\end{eqnarray}
By substituting eq.~(\ref{f_balance_end}) into eq.~(\ref{mass_cons_end}) and using eq.~(\ref{boundary2}) and (\ref{boundary_end}), one obtains the relation between $v(t)$ and $N(t)$;
\begin{eqnarray}
\left(\frac{f}{k_BT}\right)^{2 \nu -1} \left(\frac{k_BT}{\eta v(t)}\right)^{\nu} \simeq b N(t)^{\nu}
\label{N-v_1}
\end{eqnarray}
The steady state velocity $v_{ss}$ after the arrival of the tensile force at the other end is found by setting $N(t) = N_0$ in this equation;
\begin{eqnarray}
v_{ss} \simeq \frac{k_BT}{\eta_s R_0^2}({\tilde f}_{R_0})^{\frac{2\nu-1}{\nu}}
\end{eqnarray}

On the other hand, there is another relation between $v(t)$ and $N(t)$, which is available from eq.~(\ref{R_N_statistical}) and (\ref{boundary_end});
\begin{eqnarray}
b N(t)^{\nu} \simeq \frac{f}{\eta_s v(t)} -l(t) \\
=\frac{f}{\eta_s v(t)} -\int_{0}^{t}v(t') dt'
\label{N-v_2}
\end{eqnarray}
Combining eq.~(\ref{N-v_1}) and (\ref{N-v_2}) leads to an integral equation for $v(t)$, solution of which is
\begin{eqnarray}
\frac{f}{\eta_s v(t)^2} \left[ 1- B_0 \left(\frac{\eta_s k_BT v(t)}{f^2}\right)^{1-\nu}\right] \simeq t
\label{v_t}
\end{eqnarray}
(where we dropped a small term associated with the initial condition).
The time $\tau_1$ for attaining the steady state is found by setting $v(t)=v_{ss}$ in this equation;
\begin{eqnarray}
\tau_1 \simeq \tau_Z ({\tilde f}_{R_0})^{\frac{2-3\nu}{\nu}}\left[ 1-B_0 ({\tilde f}_{R_0})^{\frac{\nu-1}{\nu}}\right]
\end{eqnarray}
After rescaling the time and velocity by $\tau_1$ and $v_{ss}$ respectively, eq.~(\ref{v_t}) is rewritten as
\begin{eqnarray}
\left( \frac{v(t)}{v_{ss}}\right)^{-2} \left[ 1-B_0 ({\tilde f}_{R_0})^{-\frac{1}{\nu}}\frac{v(t)}{v_{ss}}\right] \simeq \frac{t}{\tau_1}
\end{eqnarray}

Above argument is valid when the applied force does not exceed the threshold ${\tilde f}_{b}<1$.
For larger forces, the nonlinear effect associated with the finite chain extensibility becomes apparent.
In the limit of strong forcing ${\tilde f}_{b}>N_0$, the lateral chain size of the tensed portion becomes just $y(x)=b$ and the analysis becomes very easy as in the case of absorption dynamics.
Repeating the same argument as above, one finds
\begin{eqnarray}
v_{ss} \simeq \frac{f}{\eta_s b N_0} = \frac{k_BT}{\eta_s b^2}\frac{{\tilde f}_b}{N_0}
\end{eqnarray}
\begin{eqnarray}
\tau_1 \simeq \frac{\eta_s b^2 N_0^2}{f} = \tau_R ({\tilde f}_{R_0})^{-2}{\tilde f}_b N_0
\label{tau_1_end}
\end{eqnarray}
\begin{eqnarray}
\left( \frac{v_{ss}}{v(t)}\right)^2\left[ 1- B_0 \left( \frac{v_{ss} N_0}{v(t)}\right)^{\nu-1}\right]\simeq \frac{t}{\tau_1}
\end{eqnarray}
The asymptotic form of the velocity $v(t)$ of the pulled end is
\begin{eqnarray}
v(t) \simeq \left( \frac{f}{\eta_s t}\right)^{\frac{1}{2}}
\end{eqnarray}
In the intermediate case $1<{\tilde f}_b <N_0$, one should have crossover between above two regimes just like the absorption dynamics.
Equation~(\ref{tau_1_end}) was proposed and confirmed by Monte Carlo simulation in the limit of strong forcing~\cite{Kantor_Kardar}.
It is important to notice that in the case of pulling one end, too, the earlier conjecture is approved for the strong force limit only.

\end{document}